\documentclass{rspublic}
\usepackage{amsmath,amssymb,amsfonts}
\usepackage{amsthm}
\usepackage{harvard}

\newcommand{\ket}[1]{\ensuremath{| #1 \rangle}}

\newcommand{\kb}[1]{| #1  \rangle\langle #1|}

\DeclareMathOperator{\tr}{tr}

\newcommand{\ZZ}{\mathbb{Z}}

\begin{document}
\title{Monogamy of nonlocal quantum correlations}

\author{Ben Toner}
\affiliation{CWI, Kruislaan 413, 1098 SJ Amsterdam, The Netherlands\\Institute for Quantum Information, California Institute of
Technology, Pasadena, CA 91125, USA \\(bentoner@bentoner.com)}
\date{\today}

\maketitle

\begin{abstract}{quantum, nonlocality, Bell inequality, interactive proof systems}
  We describe a new technique for obtaining Tsirelson bounds, or upper
  bounds on the quantum value of a Bell inequality.  Since quantum
  correlations do not allow signaling, we obtain a Tsirelson bound by
  maximizing over all no-signaling probability distributions.  This
  maximization can be cast as a linear program.  In a setting where
  three parties, A, B, and C, share an entangled quantum state of
  arbitrary dimension, we: (i) bound the trade-off between AB's and
  AC's violation of the CHSH inequality, and (ii) demonstrate that
  forcing B and C to be classically correlated prevents A and B from
  violating certain Bell inequalities, relevant for interactive proof
  systems and cryptography.
\end{abstract}

\section{Introduction}
One of the remarkable properties of quantum entanglement is that it is
monogamous: if Alice (A), Bob (B), and Charlie (C) each have a qubit,
and A and B are maximally entangled, then C's qubit must be completely
uncorrelated with either A's or B's.  This property is inherently
nonclassical: if A, B, and C have bits instead of qubits, and A's bit
is always the same as B's bit, then there is no restriction on how A's
bit is correlated with C's bit.  In this work, we consider the
correlations that result from making local measurements on a
multipartite quantum system.  Some such quantum correlations violate
Bell inequalities~\cite{Bell:64a}.  We show how these 
correlations,
termed {\it nonlocal}, can also be monogamous.

Consider, for example, the well-known CHSH inequality~\cite{Clauser:69a}.  Two parties, A and B, share a
quantum state $\rho$, and each chooses one of two observables to
measure on their component of the state.  Define the {\it CHSH
  operator}
\begin{align}
\label{eq:1}
  {\cal B}_\text{CHSH} = {\bf A}_1 \otimes \left({\bf B}_1 + {\bf B}_2\right) +
  {\bf A}_2 \otimes \left({\bf B}_1 - {\bf B}_2\right).
\end{align}
where ${\bf A}_1$ and ${\bf A}_2$ (${\bf B}_1$ and ${\bf B}_2$) are
A's (B's) observables and are Hermitian operators with spectrum in $[-1,+1]$.  Then the CHSH inequality states that $
\left|\left\langle {\cal B}_\text{CHSH} \right\rangle_\text{LHV}
\right| \leq 2 $, for all local hidden variable (LHV) models. There are, however,
 observables on an \emph{entangled} state, e.g., the singlet state of two qubits $\ket
{\psi^-} = \left(\ket {01} - \ket {10}\right)/\sqrt 2$, such that
$\left\langle {\cal B}_\text{CHSH} \right\rangle_\text{QM} = \tr
\left( {\cal B}_\text{CHSH}\kb{\psi^-}\right) = 2\sqrt2 $.  Thus the
correlations arising from measuring this state cannot be described by any LHV model.  In fact, it is true
that $ \left|\tr \left( {\cal B}_\text{CHSH}\rho\right) \right| \leq
2\sqrt2 $ for all observables ${\bf A}_1$, ${\bf A}_2$, ${\bf B}_1$,
${\bf B}_2$, and all states $\rho$.  Such a bound on the maximum
entangled value of a Bell inequality is termed a Tsirelson
bound~\cite{Tsirelson:80a}.  Although we do not yet know how to calculate
the best such bound for an arbitrary Bell inequality, a number of ad
hoc techniques have been
developed~\cite{Tsirelson:80a,Tsirelson:85b,Cleve:04a,Buhrman:04a,masanes05:_extrem,navascues:_bound,navascues08:long,doherty:_bounds}.

In this paper, we introduce a new technique for obtaining Tsirelson
bounds.  Since local measurements on spatially-separated components of
a multipartite quantum system can be carried out simultaneously, such
measurements cannot be used to send a signal from one party to
another.  The outcomes of local measurements on an entangled quantum
state are therefore described by a {\it no-signaling\/} probability
distribution.  Maximization over no-signaling probability
distributions can be cast as a linear program, and so we obtain an
upper bound by solving this linear program. This gives an efficient
algorithm for obtaining Tsirelson bounds. 

We use this new technique to
study the monogamy of quantum and no-signaling correlations. Suppose three parties, A, B, and C, share an entangled
quantum state of arbitrary dimension. We start by
bounding the trade-off between (A and B) and (A and C)'s violation of the
CHSH inequality. Thus we prove an analogue for quantum correlations of the famous theorem of~\citeasnoun{Coffman:00a}, which describes the tradeoff
between how entangled A is with B, and how entangled A is with C. 

In our second application, we illustrate a way to \emph{prevent} A and
B from violating a Bell inequality, even if they can share
an arbitrary entangled state. We do this by introducing an extra party C, and
forcing one of the parties, say B, to be classically correlated with C. For the Odd Cycle
Bell inequality of \citeasnoun{Cleve:04a}, we prove that the presence
of C prevents A and B from violating the Bell
inequality \emph{at all}. Finding methods to prevent entangled parties
from violating Bell inequalities is important because it allows us to
extend results about the computational hardness of computing the
classical value of a Bell inequality to the quantum domain. Indeed,
subsequent to the work described in this paper, \citeasnoun{kempe:immunization} have used the same
idea (but entirely different proof techniques) to show that it is
NP-hard to calculate, or even to approximate to exponential precision,
the entangled value of a 3-player Bell
inequality.

Our second result also has a cryptographic interpretation.  Suppose that
A and B are trying to share a secret key, and that C is 
eavesdropping on them.  If A and B observe correlations that would
cause them to win the 2-player odd cycle game with probability
greater than $1- 1/2n$, then this limits how correlated C
can be with B. \citeasnoun{Barrett:05a} have presented a key
distribution protocol along these lines, which is provably secure against no-signaling
eavesdroppers \citeaffixed{acin05:_from_theor_secur_quant_key_distr}{see also}.

The remainder of this paper is structured as follows. After defining our framework in Section~\ref{sec:framework}, we describe the linear program in Section~\ref{sec:linear}. We then present our applications: in Section~\ref{sec:ckw} we study monogamy of the CHSH inequality, while in Section~\ref{sec:oddcycle} we study the Odd Cycle Bell inequality. 

\section{Framework}\label{sec:framework}
\subsection{Nonlocal games} We cast our results in the language of {\it
  nonlocal games}, also known as {\it
  cooperative games of incomplete information} \citeaffixed{Cleve:04a}{see}.  Let $V: \ZZ_2^{m}\times \ZZ_n^{m}
\to [0,1]$ be a function and let $\pi$ be a probability distribution on $\ZZ_n^m$.
Then $V$ and $\pi$ define a $m$-player nonlocal game $G(V,\pi)$ as
follows:  A referee chooses a set of questions
$(q_1,q_2,\ldots,q_m)\in \ZZ_n^m$ randomly, according to $\pi$, and
sends question $q_i$ to player $i$.  Each player must answer with a
bit $a_i$.  The players are not permitted to communicate after
receiving the questions, but they may agree on a 
strategy before receiving them.  They win with
probability $V(a_1,a_2,\ldots,a_m|q_1,q_2,\ldots,q_m)$ (where the $|$
in $V(\cdot | \cdot)$  separates answers from questions).  The {\it
  classical value\/} of a game $G(V,\pi)$, denoted $\omega_c(G)$, is the
maximum probability with which the players can win, assuming
they use purely classical strategies.  The {\it quantum value},
denoted $\omega_q(G)$, is the maximum winning probability, assuming
they are allowed to share an arbitrary entangled state.  The {\it no-signaling
  value}, denoted $\omega_{ns}(G)$, is the maximum winning
probability, assuming the players are allowed (black box) access to
any no-signaling probability distribution.  It is clear that
$\omega_c(G) \leq \omega_q(G) \leq \omega_{ns}(G)$.

\subsection{The CHSH game} We describe how to interpret the CHSH
inequality within this framework.  The CHSH game $G_\text{CHSH}$ is
defined by setting $n=2$, letting $\pi$ be the uniform distribution on
$\ZZ_2 \times \ZZ_2$ and letting $V(a_1,a_2|q_1,q_2) = \left[a_1
  \oplus a_2 = q_1 \wedge q_2\right]$, where $a_1 \oplus a_2$ is the
exclusive-or of bits $a_1$ and $a_2$, $q_1 \wedge q_2$ is the and of bits $q_1$
and $q_2$, and $[\phi]$ is $1$ if $\phi$ is true and $0$ otherwise.
Then the winning probability of a particular strategy is $1/2 + \left
  \langle {\cal B}_\text{CHSH} \right\rangle/8$, where ${\cal
  B}_\text{CHSH}$ is the CHSH operator of Eq.~(\ref{eq:1}) and $\left
  \langle\, \cdot\, \right \rangle$ is the appropriate expectation
value for the strategy, classical or quantum. It follows
that $\omega_c(G_\text{CHSH}) = 3/4$ and $\omega_q(G_\text{CHSH}) = 1/2 +
1/(2\sqrt2) \approx 0.85$. 

\subsection{Classical correlation restricts Bell inequality violation}
Consider a 2-player game $G(V, \pi)$ that is played by A and B$_0$. Here we review work of \citeasnoun{masanes05:_gener_nonsig_theor}, who show how forcing 
B$_0$ to be {\it classically\/} correlated with additional
players B$_1$, B$_2$, \ldots, and B$_N$ restricts the advantage that A and
$B_0$ can gain by sharing entanglement.  

For $N\geq 1$, define the $N$th {\it
  extension\/} of a 2 player game $G(V,\pi)$ to be the $N+2$ player
game $G_{N}(V_N,\pi_N)$, with $\pi_N$ defined by choosing $(q_1,q_2)$
according to $\pi$ and setting $q_2 = q_3 = q_4 = \cdots = q_{N+2}$;
and
\begin{align}
V_{N}(\{a_i\}|\{q_i\}) = V(a_1,
a_2|q_1,q_2)\times[a_2=a_3=a_4=\cdots=a_{N+2}].    
\end{align}
We also set $G_0 =
G$.  The idea is that we send B$_0$'s question to the other B$_i$ and the
players  win if (i) the answers A and B$_0$ give satisfy the
winning condition of $G(V,\pi)$ and (ii) all the B$_i$ agree.

\begin{theorem}[Result~1 of \citeasnoun{masanes05:_gener_nonsig_theor}]\label{theorem:2}
Let $G(V, \pi)$ be a two player game.  Then the
values of its extensions satisfy
\begin{enumerate}
\item[(i)] $\omega_c(G_N) = \omega_c(G)$ for
all $N$, and 
\item[(ii)] $\omega_{ns}(G_N)$ is a nonincreasing sequence in $N$,
with $\omega_{ns}(G_{n-1}) = \omega_c(G)$, where $n$ is the number of
questions for B$_0$ in G.  
\end{enumerate}
\end{theorem}


 \citeasnoun{Terhal:03a}, building on work of \citeasnoun{werner89:_applic_inequal_quant_state_exten_probl}, earlier proved a similar result for $\omega_q(G_N)$.

\section{Tsirelson bounds by linear programming}\label{sec:linear}
A $m$-party no-signaling probability distribution is a
set of probabilities \begin{align}p(\{a_i\}_{i=1}^m|\{q_i\}_{i=1}^m),\end{align} 
subject to 
\begin{enumerate}
\item[(i)] {\it Positivity\/}: for all $\{a_i\}$ and all $\{q_i\}$, $p(\{a_i\}|\{q_i\})\geq 0$;
\item[(ii)] {\it Normalization\/}: for all $\{q_i\}$, $\sum_{\{a_i\}} p(\{a_i\}|\{q_i\})
= 1;$
\item[(iii)] {\it No-signaling\/}: For each subset $S \subset \ZZ_m$ of the $m$
  parties, the marginal probability distribution on $\ZZ_m - S$ must be
  independent of the inputs of the parties in $S$.  In particular,
$\sum_{\{a_i:i \in S\}} p(\{a_i\}|\{q_i\})$ 
must be independent of $\{q_i: i\in S\}$ for all $\{a_i : i \not\in
  S\}$ and for all $\{q_i : i \not\in S\}$.
\end{enumerate}
The no-signaling value of $G$ is given by 
\begin{equation*}
  \omega_{ns}(G) = \max_p \sum_{\{a_i\},\{q_i\}}\pi(\{q_i\}) V(\{a_i\}|\{q_i\}) p(\{a_i\}|\{q_i\}),
\end{equation*}
subject to the three sets of linear constraints enumerated above.  We
observe that $\omega_{ns}(G)$ is the solution to a linear program in
variables $p(\{a_i\}|\{q_i\})$.  Solving this program for
$\omega_{ns}(G)$ gives an upper bound on $\omega_q(G)$.  Moreover,
even if we cannot solve the linear program, we can obtain an upper
bound on $\omega_{ns}(G)$ by constructing a solution to the dual
program~(see \citeasnoun{Boyd:04} for an introduction to convex
optimization).

Note that for the
CHSH game, the no-signaling value $\omega_{ns}(G_\text{CHSH}) = 1$~\cite{Popescu:94a}, so the linear-programming technique provides only a trivial bound on $\omega_q(G_\text{CHSH})$.

\section{An analogue of the CKW theorem for nonlocal quantum correlations}\label{sec:ckw}
Suppose three parties, A, B,
and C, each have a qubit.  There is a well known theorem
of \citeasnoun{Coffman:00a} that describes the tradeoff
between how entangled A is with B, and how entangled A is with C.  It states that
${\cal C}_{AB}^2 + {\cal C}_{AC}^2 \leq 4 \det \rho_A$,
where ${\cal C}_{AB}$ is the concurrence between A and B, ${\cal
C}_{AC}$ is the concurrence between A and C, and $\rho_A$ is the
reduced density matrix of A.
  
To derive a similar expression for correlations, we consider a
generalization of the CHSH game to three players, suggested by Michael
Nielsen \citeaffixed{scarani01:_quant_commun_n_partn_inequal}{see also}.  In the new game, $G'_\text{CHSH}$,
the referee sends bits chosen uniformly at random to each of the three
players, and with probability $1/2$ checks if $a_1 \oplus a_2 = q_1
\wedge q_2$ and with probability $1/2$ checks if $a_1 \oplus a_3 = q_1
\wedge q_3$.  Formally, $\pi$ is uniform on $\ZZ_2^3$ and
\begin{align}V(a_1,a_2,a_3|q_1,q_2,q_3) = \frac12 [a_1 \oplus a_2 = q_1 \wedge q_2] +
\frac12 [a_1 \oplus a_3 = q_1 \wedge q_3].\end{align}  Then the winning probability of
a particular strategy is \begin{align}\frac12 + \frac1{16}\left \langle {\cal
    B}_\text{CHSH}^\text{AB} \right\rangle + \frac1{16}\left \langle {\cal
    B}_\text{CHSH}^\text{AC} \right\rangle,\end{align} where the superscripts
denote on which parties the CHSH operator acts.  It is easy to see
that $\omega_c(G'_\text{CHSH}) = 3/4$ (a strategy where everyone
always answers 0 achieves this, and this strategy is the best
possible, by the CHSH inequality applied to AB and BC separately).  It
turns out that $\omega_{ns}(G'_\text{CHSH}) = 3/4$ too, as is easily
verified using linear programming software, which implies:

\begin{theorem}\label{theorem:1}
Suppose three parties, A, B, and C share any quantum state $\rho$ (of
arbitrary dimension) and each chooses to measure one of two observables.  Then 
\begin{align}\label{eq:2}
\left|\tr \left ( {\cal B}_\text{CHSH}^\text{AB} \rho
\right)\right| + \left|\tr \left ( {\cal B}_\text{CHSH}^\text{AC} \rho
\right)\right| \leq 4.
\end{align}
\end{theorem}

Theorem~\ref{theorem:1} establishes a tradeoff between AB's and AC's violation of
the CHSH inequality.  In particular, CHSH correlations are monogamous:
if AB violate the CHSH inequality, then AC cannot, as has already been shown
for no-signaling correlations by~\citeasnoun{masanes05:_gener_nonsig_theor}.
Note that if AB and AC each share an EPR pair,
there are measurements such that either $\tr \left ( {\cal
    B}_\text{CHSH}^\text{AB} \rho \right)$ or $\tr \left ( {\cal
    B}_\text{CHSH}^\text{AC} \rho \right)$ is $2\sqrt 2$, which at
first appears to contradict Theorem~\ref{theorem:1}.  It does not: in Theorem~\ref{theorem:1} we
insist that A's observables are the {\it same\/} in $\tr \left ( {\cal
    B}_\text{CHSH}^\text{AB} \rho \right)$ and $\tr \left ( {\cal
    B}_\text{CHSH}^\text{AC} \rho \right)$.  This is in the same
spirit as the requirement of CKW that B and C are entangled with the
{same} qubit of A.
It is straightforward to generalize this result:

\begin{corollary}
Suppose $N+2$ parties A, B$_0$, B$_1$,\ldots,
B$_{N}$ share a quantum state and each chooses to measure one of two
observables.  Then A violates the CHSH inequality with at most one of
the B$_i$.  
\end{corollary}

\begin{proof}
Suppose A violates the CHSH inequality with both B$_j$ and
B$_k$, $j \neq k$.  Trace out the rest of the B$_i$'s.  We obtain a contradiction
with Theorem~\ref{theorem:1}.  
\end{proof}

For no-signaling probability distributions, we also have a converse of
Theorem~\ref{theorem:1}: for any pair $\left(\left\langle{\cal
      B}_\text{CHSH}^\text{AB}\right\rangle, \left\langle{\cal
      B}_\text{CHSH}^\text{AC}\right\rangle\right)$ consistent with
Inequality~(\ref{eq:2}), there is a no-signaling probability
distribution with these expectation values.  This is because we can
write $\left(\left\langle{\cal B}_\text{CHSH}^\text{AB}\right\rangle,
  \left\langle{\cal B}_\text{CHSH}^\text{AC}\right\rangle\right)$ as a
convex combination of $(4,0)$, $(0,4)$ and $(0,0)$, each of which is
achieved by a no-signaling probability distribution.  Thus
Inequality~(\ref{eq:2}) establishes precisely which values of
$\left(\left\langle{\cal B}_\text{CHSH}^\text{AB}\right\rangle,
  \left\langle{\cal B}_\text{CHSH}^\text{AC}\right\rangle\right)$ are
allowed.  For quantum
theory, \citeasnoun{toner:circle} have subsequently shown that the allowed region is described by 
$\left\langle{\cal B}_\text{CHSH}^\text{AB}\right\rangle^2
 + \left\langle{\cal
       B}_\text{CHSH}^\text{AC}\right\rangle^2 \leq 8$.

\section{The Odd Cycle Game}\label{sec:oddcycle} Our second example illustrates how violation of a Bell
inequality can preclude classical correlation with
another party.  We start with a 2 player game based on an interactive
proof for graph colorability.  Imagine that the two players, A and B,
are trying to convince the referee that an odd cycle of length $n$ is
2-colorable (which it is not, as $n$ is odd).  The referee sends them
each the name of a vertex, such that the two vertices are either the
same or adjacent.  A and B each send one of two colors back to the
referee.  The referee requires that, when the vertices are the
same, the two colors should agree and, when the vertices are adjacent,
the colors should differ.  Formally, we define a 2-player game
$G_\text{OC}$ as follows: Let $n\geq 3$ be an odd integer, let $\pi$
be uniform over the set $\{(q_1,q_2) \in \ZZ_n \times
\ZZ_n\,:\,\mbox{$q_1=q_2$ or $q_1+1\equiv q_2\:(\mbox{mod}\,n)$}\}$
and let $V$ be defined by $V(a_1,a_2|q,q) = [a_1 = a_2]$,
$V(a_1,a_2|q,q+1) = [a_1 \neq a_2]$.  It is established by \citeasnoun{braunstein90:_wring} and \citeasnoun{Cleve:04a} that $\omega_c(G_\text{OC}) = 1 - 1/2n$ and
$\omega_q(G_\text{OC}) = \cos^2(\pi/4n)$. The Odd Cycle Game has also been considered by \citeasnoun{10.1109/CCC.2007.39}, who study its behaviour under parallel repetition.

Now consider the first extension of this game, which we denote $G'_\text{OC}$. Formally,
$G'_\text{OC}$ is defined by using the same distribution as
$G_\text{OC}$ on $(q_1, q_2)$, and setting $q_2=q_3$.  The function
$V$ is defined by
\begin{align}
V(a_1,a_2,a_3|q,q,q) &= [a_1 = a_2=a_3],\\
V(a_1,a_2,a_3|q,q+1,q+1) &= [a_1 \neq a_2 = a_2].
\end{align}
Theorem~\ref{theorem:2} implies
that $\omega_c(G'_\text{OC}) = \omega_c(G_\text{OC}) = 1-1/2n$.  We
shall show:
\begin{theorem} \label{theorem:3}
For the first extension of the odd cycle game, 
$\omega_c(G'_\text{OC}) =
\omega_{q}(G'_\text{OC}) = \omega_{ns}(G'_\text{OC}) = 1 -
1/2n$.
\end{theorem}
Thus sharing entanglement (or indeed no-signaling correlations) gives
no advantage for $G'_\text{OC}$.  This
result is remarkable because it establishes that adding just one
additional player is sufficient to prevent A and B from gaining
advantage by sharing entanglement, rather than the $n-1$ additional
players required in Theorem~\ref{theorem:2}.
In the context of interactive proof
systems, we can interpret the fact that $\omega_q(G_\text{OC}) >
\omega_c(G_\text{OC})$ in the 2-player game as saying that sharing
entanglement allows the provers to cheat, because it increases the
probability with which they are able to convince the referee that the
odd cycle is 2-colorable.  Theorem~\ref{theorem:3} shows that we can counter this by
adding an extra prover, and forcing B to be classically correlated
with her.  This placed no extra burden on classical provers, because
an optimal classical strategy is deterministic, but it prevents
quantum provers from gaining any advantage by sharing entanglement.

\begin{proof}[Proof of Theorem~\ref{theorem:3}]
There are a number of symmetries we can use to simplify the problem.
Without changing the probability of winning: 
\begin{enumerate}
\item[(i)] all parties can flip
their outputs, and/or
\item[(ii)] all parties can add ($\text{mod}\, n$) an
integer $m$ to their inputs, and/or 
\item[(iii)] B and C can exchange roles.
\end{enumerate}
For
a given no-signaling strategy, let $p(a,b,c|i,j,k)$ be the probability
that (A,B,C) answer $(a,b,c)$ when asked $(i,j,k)$.
Then we can take $p(a,b,c|i,j,k)$ to be symmetric under these three
symmetries.  In particular, symmetry (i) implies we can
restrict attention to $a=0$, symmetry (ii) to $i=0$. Therefore, let 
$r(b,c|j,k) = p(0,b,c|0,j,k)$.
We shall use symmetry (iii) to give extra constraints, rather than to
reduce the number of parameters.  

We rewrite the primary linear program in these variables, labeling the
constraints.  Our goal is to maximize 
\begin{align}
  \label{eq:5}
  \omega_{ns}(G'_\text{OC}) = \frac{1}{2} \max_r \left[ r(0,0|0,0) + r(1,1|1,1) \right],
\end{align}
subject to
\begin{itemize}
\item (Normalization)  $n(j,k)$:
  \begin{align}
    \sum_{b,c} r(b,c|j,k) = 1,
  \end{align}
for $0 \leq j,k < n$.
\item (Symmetry)  $s(b,c|j,k)$:
  \begin{align}
r(b,c|j,k) = r(c,b|k,j),
  \end{align}
for $b,c\in \{0,1\}$, $0 \leq j,k < n$.  Note that when $b=c$ and $j=k$
this constraint is trivial.
\item (No-signaling conditions, A to BC) $y(d|j,k)$:
  \begin{align}
p(0,d|j,j+k) + p(1,\bar d|j,j+k)  = p(0,d|0,k) + p(1,\bar d|0,k),    
  \end{align}
for $d \in \{0,1\}$, $1 \leq j < n$, $0 \leq k < n$, where the sum $j+k$ is taken mod $n$.  
\item (No-signaling conditions, B to AC) $z(d|j,k)$:
  \begin{align}
p(0,d|j,k) + p(1,d|j,k) = p(0,d|0,k) + p(1,d|0,k),    
  \end{align}
for $d \in \{0,1\}$, $1 \leq j < n$, $0 \leq k < n$, where the sum $j+k$ is taken mod $n$.  
\end{itemize}
We omit the no-signaling conditions in the other directions
(BC to A and AC to B), which do not further constrain the solution.

Each constraint in the
primary linear program corresponds to a variable in the dual, as
labeled above.  The objective of the dual program is to 
minimize 
\begin{align}
  \label{eq:objdual}
  \frac{1}{2 n} \sum_{j,k} n(j,k).
\end{align}
subject to the constraints $\mu(0,0|0,0), \mu(1,1|1,1) \geq n$, 
$\mu(b,c|j,k) \geq 0$, for all $b, c \in \{0,1\}$, $0 \leq  j, k < n$, where 
\begin{align}
\mu(b,c|j,k) &= n(j,k) + s(b,c|j,k) - s(c,b|k,j)
\nonumber \\&\qquad+\left[j=0\right]\sum_{j'=1}^{n-1}\left( y \left ( \frac{1-bc}{2}\bigg|j',k\right)
+ z(c|j',k)\right)
\nonumber
\\&\qquad\qquad-\left[j\neq 0\right]\left(y\left(\frac{1-bc}{2}\bigg|j,k-j\right)+z(c|j,k)\right).
\end{align}

We now give an explicit solution to the dual.  This
solution was constructed by numerically solving the linear program for small
$n$, iteratively converting the inequality constraints into a consistent set of equality conditions by
generalizing from the small $n$ solutions, and then inverting these
constraints to yield the solution. While we used numerical methods to obtain the solution, we emphasize that the final solution can be checked analytically, without any need for numerical methods. The nonzero variables are: 
\begin{align}
n(0,0) &= 2n - 1;&\\
s(0,1|0,0) &= 3n/2,&\\
s(0,1|1,0) &= -n + 1,&\\
s(0,0|0,1) &= -n+1,&\\
s(0,1|1,1) &= -n/2,&\\
s(0,0|j, j+1) &= {(-1)^j} &\text{ for } j =
1, 2, \ldots, n-1;\\
s(0,1|j, j+1) &= -{(-1)^j} &\text{ for } j =
1, 2, \ldots, n-1;\\
y(0|1,0) &= {-2n+3},&\\
y(0|1,k) &= -n + k +5/2+(-1)^{k}/2 &\text{ for } k =
1, 2, \ldots, n-1;\\
y(1|1,0) &= 3 - 3n/2,&\\
y(1|1,1) &= {-n+4},&\\
y(1|j,1) &= -{(-1)^j}&\text{ for } j = 2,3, \ldots, n-1,\\
y(1|1,k) &= -n+k+5/2 +(-1)^{k}/2 &\text{ for } k = 2, 3, \ldots, n-2,\\
y(1|1,n-1) &= {-n+3},&\\
y(1|j,n-1) &= 1 - (-1)^j&\text{ for } j = 2,3, \ldots, n-1;\\
z(0|1,0) &= {n-3},&\\
z(0|1,1) &= {2n-3},&\\
z(0|1,2) &= {n-4},&\\
z(0|j, j-1) &= - {1} &\text{ for } j = 2, 3, \ldots, n-1,\\
z(0|j, j+1) &= {(-1)^j} &\text{ for } j = 2, 3, \ldots, n-1,\\
z(0|1,k) &= n-k -3/2 + (-1)^{k}/{2}
&\text{ for } k = 3, 4, \ldots, n-1;\\
z(1|j, j-1) &= -1 + (-1)^j &\text{ for } j = 1, 2, \ldots,n-1,\\
z(1|1,k) &= n-k-3/2  +  (-1)^{k}/{2} &\text{ for } k = 1, 2, \ldots, n-1.
\end{align}
All other variables are zero.  

For this solution, it's tedious but straightforward
to establish that: 
$\mu(0,0|0,0) = \mu(1,1|1,1) = n$,
$\mu(0,1|0,0) = 2n$,
$\mu(1,0|0,1) = 2n-2$,
$\mu(0,0|j,j-1) = 1 + (-1)^j${ for } $j = 1, 2,\ldots, n-1$,
$\mu(1,1|k+1,k) = 1+(-1)^k${ for } $k = 1,2,\ldots, n-1$, and 
$\mu(b,c|j,k) = 0$,{ otherwise}.
Thus our solution satisfies the constraints.  Substituting into
Eq.~(\ref{eq:objdual}), we find that $\omega_{ns}(G'_\text{OC}) \leq 1
- 1/2n$, which proves Theorem~\ref{theorem:3}.
\end{proof}

\section{Acknowledgements} I thank Richard Cleve, John Preskill,
Graeme Smith, Frank Verstraete, and John Watrous for useful
suggestions, and Michael Nielsen for suggesting Theorem~\ref{theorem:1}.  This work
was supported in part by the NSF under grant PHY-456720, the ARO under
Grant No. W911NF-05-1-0294, the EU FP6-FET Integrated Project QAP
CT-015848, NWO VICI project 639-023-302, and the Dutch BSIK/BRICKS
project.


\end{document}